\documentclass[conference]{IEEEtran}
\IEEEoverridecommandlockouts
\usepackage{cite}
\usepackage{amsmath,amssymb}
\usepackage{graphicx}
\usepackage{booktabs}
\usepackage{multirow}
\usepackage{makecell}
\usepackage{xcolor}
\usepackage{tikz}
\usetikzlibrary{arrows.meta,positioning}
\usepackage[hidelinks]{hyperref}

\begin{document}

\title{IoMT-SecAlarmBench: A Semi-Synthetic Counterfactual Benchmark for Integrity Attacks in IoMT}

\author{\IEEEauthorblockN{Emmanuel C. Ugwuabonyi}
\IEEEauthorblockA{\textit{Dept. of Computer Science and Electrical Engineering}\\
\textit{University of Maryland, Baltimore County}\\
Baltimore, MD, USA\\
ugwuabonyi@umbc.edu}
\and
\IEEEauthorblockN{Dmitri Perkins}
\IEEEauthorblockA{\textit{Dept. of Computer Science and Electrical Engineering}\\
\textit{University of Maryland, Baltimore County}\\
Baltimore, MD, USA\\
dmitrip1@umbc.edu}}

\maketitle

\begin{abstract}
Networked bedside monitors can be attacked anywhere along the path from the sensor to the patient record, and a falsified measurement can closely resemble the sensor artifacts and genuine physiological deteriorations that already account for most ICU alarms. We present IoMT-SecAlarmBench, a semi-synthetic benchmark that injects controlled integrity attacks into authentic ECG+PPG recordings from 627 records in the PhysioNet/CinC 2015 dataset. A factorial design combines four attack shapes, four severity levels, and two plausibility regimes, includes replay attacks, and repeats the process across three attacker capability levels: modifying a single device stream, altering both channels independently, and altering both channels consistently. Because every attack is injected, the original clean waveform is known. This enables counterfactual evaluation at two levels: the waveform itself and the decision of a specified rate-alarm algorithm. Six unsupervised detectors struggle with replay attacks and short spikes, and a dedicated heart-rate and pulse-transit coherence baseline does not resolve these failures. Joint ECG+PPG replay and coordinated heart-rate warping perform at chance against their matched clean hosts (host-matched AUROC 0.49--0.55). At the alarm level, a monitor that requires ECG and PPG agreement loses 38\% of the rate alarms raised on true-alarm records when only the PPG stream is replayed, and 48\% under coordinated replay attacks that evade all detectors. A supervised classifier separates attacks effectively (F1 = 0.97) but almost never identifies genuine physiological events, even when attacks and controls are matched by position (event F1 up to 0.02). We release the code, attack generators, and benchmark datasheet.
\end{abstract}


\section{Introduction}

Connected medical devices are being deployed faster than the systems designed to secure them, while the signals they produce remain inherently noisy. Drift, motion artifacts, calibration errors, and patient-to-patient variability drive ICU false-alarm rates (FAR) as high as 86\%~\cite{clifford2015}. When an abnormal measurement triggers an alarm, an anomaly detector may determine that the reading is unusual, but it cannot identify whether the cause is genuine physiological deterioration, a drifting or detached sensor, or a malicious value injected along the measurement path. Each scenario requires a different response, and confusing them can jeopardize patient safety. The clinically meaningful question is therefore counterfactual: \emph{would the alarm have fired in the absence of the attack?} Real-world recordings cannot answer this because they contain only the observed signal, not the signal that would have been recorded otherwise. Causal inference addresses this limitation through semi-synthetic designs that simulate outcomes on real covariates, making both potential outcomes available~\cite{hill2011}. In our setting, attacks are injected into authentic data, so the corresponding clean signal is known.

Stealthy sensor attacks have been extensively studied in industrial cyber-physical systems~\cite{teixeira2012,teixeira2015,bonczek2020}, and causal root-cause methods are commonly evaluated on benchmarks with known causes, where many perform only slightly better than chance~\cite{pham2024rca}. However, the IoMT community lacks a benchmark that combines genuine coupled physiology, integrity attacks on the measurement path, labeled faults and clinical events as hard negatives, and a counterfactual reference for every attack. 
\noindent We make the following contributions:
\vspace*{-2pt}\begin{enumerate}
\item \textbf{A benchmark with counterfactual references.} 
IoMT-SecAlarmBench injects 33--38 attack conditions into each of 1,173 clean host windows derived from 627 CinC 2015 records~\cite{clifford2015}. Every attacked window retains its attack type, parameters, and the clean waveform it replaced, while a specified rate-alarm algorithm converts each clean--attacked pair into an alarm-level counterfactual.
\item \textbf{An attacker-capability axis.} 
Attacks are generated for a single stream (C1, ECG or PPG), for both channels altered independently (C2), and for both channels altered consistently (C3), all applied to the same host windows.
\item \textbf{An evaluation protocol and baselines.} 
We evaluate six detectors, a physiological-coherence baseline, and a supervised classifier using threshold-free and fixed-budget metrics, host-matched and position-matched controls, seed variance analysis, and record-grouped bootstrap confidence intervals.
\item \textbf{Audits.} A label audit of the CinC alarm zone and an independent leakage audit of WUSTL-EHMS-2020.
\end{enumerate}
The generator is deterministic and versioned. Approximately 200,000 labeled windows across the four attacker settings can be reconstructed from the public records using a manifest of random seeds and module hashes, making the benchmark a reproducible time-series dataset. Synchronized network traffic and causal attribution methods are beyond the scope of this release.

\section{Related Work}

\textbf{Datasets.} CICIoMT2024~\cite{dadkhah2024} includes forty devices and eighteen attack classes but contains no physiological signals. WUSTL-EHMS-2020~\cite{hady2020} pairs network traffic with a small set of vital-sign summaries under man-in-the-middle and injection attacks, yet provides no waveforms. MIMIC-III Waveform~\cite{moody2020} contains long ICU waveforms, while CinC 2015~\cite{clifford2015} pairs ECG and PPG signals and labels each life-threatening alarm as true or false. SWaT and WADI~\cite{mathur2016} are industrial control-system datasets without patient information, and general anomaly benchmarks such as SMAP, MSL, and SMD contain simple anomalies and known labeling issues~\cite{wu2023flawed}.

\textbf{Attack injection.} Real IoMT integrity attacks are uncommon, legally reportable, and unsafe to stage in clinical environments. As a result, controlled injection into genuine recordings has become the standard evaluation approach~\cite{hasan2026}. We adopt the four signal modifications proposed by Teixeira et al.~\cite{teixeira2012} and adapted to physiological streams by DSTAN-Med~\cite{hasan2026}. Related work has examined false-data injection~\cite{liu2009}, replay attacks~\cite{mo2009}, resource-constrained adversaries~\cite{teixeira2015}, stealthy attacks against residual-based detectors~\cite{bonczek2020}, morphological ECG manipulation~\cite{cai2016}, and medical sensor spoofing~\cite{park2016}. For dataset documentation, we follow Datasheets for Datasets~\cite{gebru2021} and DAIMS~\cite{marandi2025}.

\section{Threat Model}
\label{sec:threat}

\begin{figure}[t]
\centering
\begin{tikzpicture}[font=\scriptsize,>={Stealth[length=4pt]},
  st/.style={draw,circle,inner sep=1pt,minimum size=13pt},
  ch/.style={draw,rectangle,minimum width=22pt,minimum height=10pt,inner sep=1.5pt},
  atk/.style={draw=red!75!black,dashed,rectangle,minimum width=22pt,minimum height=10pt,inner sep=1.5pt,text=red!75!black}]
\node[st] (S1) at (0,0) {$S_t$};
\node (I1) at (0,0.75) {$\mathrm{do}(S_t{=}\tilde s)$};
\node[ch] (E1) at (-0.5,-0.9) {ECG};
\node[ch] (P1) at (0.5,-0.9) {PPG};
\draw[->,red!75!black,thick] (I1) -- (S1);
\draw[->] (S1) -- node[left,pos=0.4]{$f_E$} (E1);
\draw[->] (S1) -- node[right,pos=0.4]{$f_P$} (P1);
\node[align=center] at (0,-1.45) {(a) clinical event:\\channels move together};
\begin{scope}[xshift=2.85cm]
\node[st] (S2) at (0,0) {$S_t$};
\node (I2) at (0.55,0.75) {$\mathrm{do}(x_{P,t}{=}\tilde x)$};
\node[ch] (E2) at (-0.5,-0.9) {ECG};
\node[atk] (P2) at (0.5,-0.9) {$\tilde x_P$};
\draw[->] (S2) -- node[left,pos=0.4]{$f_E$} (E2);
\draw[->,densely dotted,gray] (S2) -- node[right,pos=0.4]{$f_P$} (P2);
\draw[->,red!75!black,thick] (0.75,0.62) to[out=-80,in=70] (P2.north east);
\node[align=center] at (0,-1.45) {(b) C1 attack: one\\channel stops tracking $S_t$};
\end{scope}
\begin{scope}[xshift=5.7cm]
\node[st] (S3) at (0,0) {$S_t$};
\node (I3) at (0,0.75) {$\mathrm{do}(x_{E,t},x_{P,t})$};
\node[atk,minimum width=18pt] (E3) at (-0.66,-0.9) {$\tilde x_E$};
\node[atk,minimum width=18pt] (P3) at (0.66,-0.9) {$\tilde x_P$};
\draw[->,densely dotted,gray] (S3) -- (E3);
\draw[->,densely dotted,gray] (S3) -- (P3);
\draw[->,red!75!black,thick] (-0.4,0.62) to[out=-110,in=80] (E3.north west);
\draw[->,red!75!black,thick] (0.4,0.62) to[out=-70,in=100] (P3.north east);
\node at (0,-0.9) {$\simeq$};
\node[align=center] at (0,-1.45) {(c) C3 attack: channels\\agree, $S_t$ untouched};
\end{scope}
\end{tikzpicture}
\caption{Modeling assumption behind the benchmark. (a) When the sensors work, a change in patient state should appear in every channel, up to sensor-specific delay and noise. (b) A single-stream attack replaces one channel, which then stops tracking the other. (c) A coordinated attacker keeps the channels consistent and so removes that cue.}
\label{fig:scm}
\end{figure}

\textbf{Modeling assumption.} Let $S_t$ denote the patient's physiological state. Each channel $c$ is observed as $x_{c,t}=f_c(S_t,\varepsilon_{c,t})$, where $f_c$ represents the sensor process and $\varepsilon_{c,t}$ its noise. We model a genuine clinical event as an intervention on the state, $\mathrm{do}(S_t=\tilde s)$, and an integrity attack as an intervention on the measurement, $\mathrm{do}(x_{c,t}=\tilde x)$, while leaving $S_t$ unchanged (Fig.~\ref{fig:scm}). Because real sensors differ in delay, sensitivity, and noise characteristics, this assumption is an idealization. Cross-channel coherence is therefore an expected relationship but is not guaranteed, and Section~\ref{sec:res} evaluates how effectively detectors exploit it.

\textbf{Attacker.} The attacker operates between a bedside sensor and the clinical record or central station, with the ability to read and modify samples in transit. We consider three capability levels:
\begin{itemize}
\item \textbf{C1, single stream.} The attacker controls one device stream, such as an independent pulse oximeter channel, while the other channel remains an uncompromised reference.
\item \textbf{C2, both channels, independent.} The attacker controls both channels, for example at a gateway that aggregates them, but manipulates each stream separately with its own onset, magnitude, and replay source.
\item \textbf{C3, both channels, coordinated.} The attacker controls both channels while preserving consistency between them. The attacker may replay the same historical ECG+PPG segment into both streams, apply matched modifications to each channel with a pulse-transit delay offset for PPG, or jointly time-scale both channels to mimic a heart-rate change.
\end{itemize}
The attacker does not control the patient's physiological state, the detector, the detector's training data, or the labels. All attacks are predefined, and adaptive attacks that query the detector are outside the scope of this work.

\section{Benchmark Construction}
\label{sec:bench}

\textbf{Core cohort.} CinC 2015~\cite{clifford2015} contains 750 ICU records sampled at 250\,Hz, each 300\,s long, half of them extending to 330\,s. Every record contains a life-threatening arrhythmia alarm at 300\,s that experts labeled as true or false. Of these, 627 records contain both ECG and PPG signals and form the ECG+PPG cohort used in this benchmark (231 true and 396 false alarms). ECG is band-pass filtered to 0.5--40\,Hz, PPG to 0.5--8\,Hz using zero-phase fourth-order Butterworth filters. Both signals are then resampled to 125\,Hz, z-score normalized within each record, and segmented into 10\,s windows with a 5\,s stride. Flat lines, clipping, and physiologically implausible segments, including instantaneous heart rates outside 30--200\,bpm, are retained and labeled. Windows with artifact fractions above 0.10 are excluded as attack hosts.

\textbf{Attack families.} We organize attacks according to how much of the original measurement remains unchanged. \emph{Replay} substitutes a previously recorded segment from the same or a different patient. Although the signal remains physiologically realistic within each channel, it is no longer current, and a single-channel replay also loses agreement with the companion channel. \emph{False data injection (FDI)} adds a crafted perturbation that remains within a plausibility interval $[l_c,u_c]=\mu_c\pm k_c\bar\sigma_c$, computed from the pre-onset portion of each window ($k_c=5$ for ECG and 4 for PPG). \emph{Spoofing} places the signal outside that interval.

\textbf{Factorial design.} FDI and spoofing are crossed with four attack morphologies~\cite{teixeira2012,hasan2026} and four ordered severity levels (Table~\ref{tab:sev}). Let $\bar\sigma_c$ denote the pre-onset standard deviation of the window and $s\in\{-1,+1\}$ a random sign. A \emph{spike} adds $s\,m\,\bar\sigma_c$ over 3--5 samples, \emph{stuck-at} freezes the channel at a selected value, \emph{drift} introduces a linear ramp up to $\delta_{\max}$, and \emph{bias} adds a constant offset. Severity levels alternately increase amplitude and duration so that the integrated perturbation approximately doubles from one level to the next. This produces 32 crafted attack conditions, plus same-patient and cross-patient replay attacks, for 34 conditions per host. In spoofing, the injected value is fixed beyond the plausibility bound, so morphology determines only onset and duration, reducing the 16 spoofing cells to 10 unique conditions.

\begin{table}[t]
\centering
\caption{Severity schedule. Magnitudes in $\bar\sigma_c$ units; durations in seconds. $\dagger$: in-bound FDI is clipped at the plausibility bound.}
\label{tab:sev}
\setlength{\tabcolsep}{3pt}
\footnotesize
\begin{tabular}{@{}lllll@{}}
\toprule
Morph. & L1 & L2 & L3 & L4 \\
\midrule
Spike & $0.5$ & $1.0$ & $2.0$ & $4.0^\dagger$ \\
Stuck-at & $s\,U(0,1)$ & $s\,U(0,3)$ & $s\,U(0,5)^\dagger$ & $s\,U(0,10)^\dagger$ \\
Drift & $1.0$ & $2.0$ & $2.0$ & $4.0^\dagger$ \\
Bias & $s\,U(0.5,1)$ & $s\,U(1,2)$ & $s\,U(2,4)^\dagger$ & $s\,U(4,8)^\dagger$ \\
\midrule
\multicolumn{5}{@{}l@{}}{Durations: spike 0.024/0.024/0.040/0.040; stuck-at 1/2/4/4;}\\
\multicolumn{5}{@{}l@{}}{drift 2/2/4/4; bias 2/4/4/4.}\\
\bottomrule
\end{tabular}
\end{table}

\textbf{Attacker settings.} C1 applies the 34 attack conditions to either PPG, matching the first benchmark release, or to ECG. C2 applies each crafted condition to both channels using independent parameter draws and adds replay attacks with different source segments for ECG and PPG (33 conditions per host). C3 includes three attack types: joint same-patient and cross-patient replay of a prior ECG+PPG segment; the 32 crafted conditions applied consistently to both channels with matched sign and magnitude, with the PPG perturbation delayed by the estimated pulse-transit lag; and a rate-warp attack that time-scales a 4\,s segment of both channels by factors of 1.05, 1.10, 1.20, or 1.35, either accelerating or slowing the apparent heart rate (38 conditions per host).

\textbf{Labels.} Each window receives a single cause label derived from the CinC expert-reviewed alarm verdict and alarm timing. Windows ending before the alarm are \emph{physiology-normal}, whereas windows extending beyond the alarm in true-alarm records are \emph{physiology-event}. Injected windows are labeled attack. The remaining windows, previously grouped as fault/artifact, are now separated into two categories. \emph{Artifact-confirmed} windows contain artifact masks covering more than 10\% of the window (4,368 windows). \emph{Monitor false alarms} are post-alarm windows from false-alarm records with minimal detected artifact (982 windows). These demonstrate monitor error without evidence of sensor failure and are therefore not treated as confirmed faults.

\textbf{Alarm zone.} In CinC, annotated events begin within the final 10\,s before the alarm (290--300\,s), although earlier unannotated events may also occur~\cite{clifford2015}. The first benchmark release labeled windows ending within this interval as physiology-normal, even when they may have contained the event. We now mark these as \emph{alarm-zone} windows, exclude them from the normal class in a sensitivity analysis (Section~\ref{sec:audit}), and provide an alternative version that relabels them according to the expert verdict.

\textbf{Counterfactual references.} Each attacked window retains the exact clean waveform it replaced, together with its channel, onset time, duration, and attack level. This serves as the signal-level counterfactual. For alarm-level counterfactuals, we apply a specified alarm algorithm $g$ to both the clean and attacked versions. Comparing $g(X_{\mathrm{clean}})$ and $g(X_{\mathrm{attack}})$ reveals whether the attack created an alarm that would otherwise not have occurred or suppressed one that would have occurred. We implement $g$ using the CinC 2015 definitions of the three rate alarms: asystole (no detected beat for at least 4\,s), extreme bradycardia (heart rate below 40\,bpm for five consecutive beats), and extreme tachycardia (heart rate above 140\,bpm for seventeen consecutive beats). Beat detections are produced by a slope-energy QRS detector on ECG and a maximum-upslope detector on PPG (Table~\ref{tab:set}). We report results for two reference monitors. $g_{\mathrm{ECG}}$ uses ECG alone, while $g_{\mathrm{fused}}$ raises an alarm only when ECG and PPG agree on the alarm type, a common strategy for reducing false alarms. These are reference monitors, not commercial systems, and do not include ventricular alarms, which require beat-morphology analysis.

\textbf{Composition.} Table~\ref{tab:set} summarizes the benchmark. The same 1,173 host windows, with up to two hosts per record (the first and last clean pre-alarm windows), are reused across all attack conditions and attacker settings. Consequently, the 39,882 C1-PPG attack windows represent 34 variants of the same 1,173 hosts, and are not independent observations. This design prevents attack samples from outnumbering events and faults by several orders of magnitude.
\begin{table}[t]
\centering
\caption{Benchmark, baseline and model settings.}
\label{tab:set}
\scriptsize
\setlength{\tabcolsep}{3pt}
\begin{tabular}{@{}p{0.25\columnwidth}p{0.71\columnwidth}@{}}
\toprule
Source & CinC 2015, 627-record ECG+PPG core cohort \\
Windows & 10\,s, 5\,s stride, $2\times1{,}250$ samples at 125\,Hz \\
Split & 436/91/100 records; record-disjoint; stratified by alarm type \\
Features & 22: ten per channel (mean, SD, peak-to-peak, RMS, mean absolute successive difference, IQR, skewness, kurtosis, dominant frequency, spectral entropy), ECG--PPG zero-lag correlation and best lag \\
Classes & normal 32,875 (1,110 in the alarm zone); event 622; artifact-confirmed 4,368; monitor false alarm 982 \\
Attacks & 1,173 hosts; C1-PPG 39,882 (replay 2,346, FDI 18,768, spoofing 18,768); C1-ECG 39,882; C2 38,709; C3 44,574 \\
\midrule
Isolation Forest & 200 trees on standardized features; seeds 0--4 \\
USAD & full-resolution input (2,500), hidden 256, latent 40, $\alpha{=}\beta{=}0.5$, lr $10^{-4}$, batch 256; 1.96\,M parameters \\
LSTM-AE & $10\times$ pooling, hidden 64, latent 32, lr $10^{-3}$, batch 256; 45\,k \\
GDN & $10\times$ pooling, window 20, embedding 32, hidden 64, lr $10^{-3}$, batch 512; 7.9\,k \\
TranAD & $10\times$ pooling, $d_{\mathrm{model}}$ 64, 4 heads, 1 layer, lr $10^{-3}$, batch 256; 143\,k \\
OmniAnomaly & $10\times$ pooling, hidden 64, latent 16, $\beta{=}1$, lr $10^{-3}$, batch 256; 31\,k \\
All deep & Adam, 40 epochs, gradient-norm clipping at 1.0, deterministic kernels, seeds 0--4 \\
Coherence & robust z-scores (training normals) of HR--PR difference, unmatched beats, PTT SD, max PTT deviation, 3\,s beat-count mismatch; score = sum of positive z \\
Beat detectors & ECG: squared slope, 150\,ms integration, $0.3\times$ 90th-percentile peak; PPG: maximum upslope, 50\,ms smoothing, same rule \\
Random forest & class-balanced weights; 300 trees (discrimination), 200 (WUSTL) \\
\bottomrule
\end{tabular}
\end{table}

\section{Evaluation Protocol}
\label{sec:eval}

\textbf{Detectors.} Isolation Forest~\cite{liu2008} serves as the classical baseline and uses all 22 engineered features, including two cross-channel features. LSTM-AE~\cite{malhotra2016} and USAD~\cite{audibert2020} reconstruct each window, with USAD incorporating adversarial training. OmniAnomaly~\cite{su2019} is a stochastic recurrent variational autoencoder scored by likelihood. GDN~\cite{deng2021} applies graph attention across channels; with only two channels, its graph structure is minimal, so it functions primarily as a forecasting model. TranAD~\cite{tuli2022} is a transformer-based model with self-conditioned reconstruction. USAD processes full-resolution windows, while the remaining deep models use $10\times$ temporal pooling. Hyperparameters (Table~\ref{tab:set}) were fixed during implementation, sized for a two-channel setting with a shared training schedule, and never tuned. As a result, the test attacks cannot influence model selection. Because all six detectors are general-purpose methods, we also include an interpretable \emph{physiological-coherence} baseline derived from the mechanism in Fig.~\ref{fig:scm}. This baseline compares ECG heart rate and PPG pulse rate, verifies that each ECG R-peak is followed by a pulse within 80--600\,ms, and measures variation in pulse-transit time (Table~\ref{tab:set}). All seven detectors are trained exclusively on physiology-normal windows.

\textbf{Splits.} Data are partitioned at the record level, and a runtime check verifies that no record appears in more than one split. CinC 2015 does not provide patient identifiers and permits up to three alarms of each type per patient~\cite{clifford2015}. Consequently, the strongest verified guarantee is record-disjointness; the same patient may still contribute records to multiple splits.

\textbf{Metrics.} All results are reported without point adjustment. We use AUPRC, together with its chance baseline equal to the positive prevalence of the evaluated subset, and AUROC, which is prevalence-independent. For fixed-budget detection, thresholds are chosen on validation normals to achieve a target FAR and then applied unchanged to the test set, where the realized FAR is reported. We also report each method's validation-set maximum-F1 threshold to illustrate threshold variability across random seeds.

\textbf{Controls.} Attacks are injected into clean pre-alarm host windows, whereas physiological events and monitor false alarms occur after the alarm. A model could therefore distinguish classes using temporal position or record identity, without relying on attack characteristics. To control for this, \emph{host-matched AUROC} compares each attacked window only against its corresponding clean host. For cause-discrimination and alarm-suppression studies, we additionally inject all four attacker settings into post-alarm event windows and monitor-false-alarm windows (up to three per record), ensuring that attacks and controls occupy comparable positions.

\textbf{Imbalance and uncertainty.} Attack prevalence is determined by benchmark design, whereas physiological events are naturally rare (622 windows overall, including 92 in the test set). We therefore emphasize prevalence-independent metrics, preserve the natural class distribution during testing, and apply rebalancing only within training folds. Users seeking more realistic attack prevalence can sample a single attacked window per host. Five random seeds capture training variability but not record-level variation. To quantify uncertainty, we report 95\% confidence intervals from 1,000 record-grouped bootstrap resamples of the test records. Each resample retains all windows from a record, including hosts, attacks, and controls, and averages performance across seeds.

\section{Results}
\label{sec:res}

Unless otherwise noted, results are reported for C1-PPG on the 100 test records, comprising 11,825 windows, of which 6,392 are attacks (prevalence 0.541). Tables~\ref{tab:rq1}--\ref{tab:cap} are based on five seeds with deterministic kernels.

\subsection{RQ1: Detection}

Overall detection performance is moderate, and the detectors perform similarly (Table~\ref{tab:rq1}). AUPRC ranges from 0.82 to 0.90, compared with a chance level of 0.54. Isolation Forest, the only learned detector with explicit ECG--PPG coupling features, achieves the highest performance, and its interval (0.886--0.916) does not overlap with those of the other detectors. Treating artifact and monitor-false-alarm windows as negatives reduces AUPRC by 0.01--0.03 for all methods, and four of the six detectors flag these windows more often than ordinary normal windows.

The max-F1 operating point is highly unstable. Across five seeds, it changes the FAR on normal windows by as much as $\pm0.234$ and the FAR on event windows by up to $\pm0.222$ (both USAD), whereas threshold-free AUPRC varies by at most $\pm0.020$.

\begin{table}[t]
\centering
\caption{Detection, C1-PPG test split (mean $\pm$ sd, five seeds). Chance AUPRC is 0.541, or 0.512 with fault windows as negatives. FARs are at each method's own validation max-F1 threshold; for the coherence baseline that threshold flags every window.}
\label{tab:rq1}
\footnotesize
\setlength{\tabcolsep}{2pt}
\begin{tabular}{@{}lccccc@{}}
\toprule
Method & AUPRC & \makecell{AUPRC\\(+fault)} & \makecell{FAR\\normal} & \makecell{FAR\\fault} & \makecell{FAR\\event}\\
\midrule
Isol.\ Forest & .901$\pm$.004 & .872$\pm$.006 & .251$\pm$.025 & .470$\pm$.027 & .302$\pm$.045\\
TranAD & .867$\pm$.020 & .854$\pm$.021 & .143$\pm$.054 & .178$\pm$.042 & .198$\pm$.111\\
LSTM-AE & .854$\pm$.000 & .841$\pm$.000 & .165$\pm$.006 & .188$\pm$.002 & .206$\pm$.011\\
OmniAnom. & .843$\pm$.007 & .831$\pm$.008 & .554$\pm$.082 & .422$\pm$.083 & .707$\pm$.030\\
GDN & .831$\pm$.002 & .813$\pm$.002 & .250$\pm$.033 & .269$\pm$.019 & .420$\pm$.054\\
USAD & .817$\pm$.014 & .801$\pm$.015 & .604$\pm$.234 & .569$\pm$.247 & .639$\pm$.222\\
Coherence & .655 & .603 & -- & -- & --\\
\bottomrule
\end{tabular}
\end{table}

\subsection{RQ2: Stealthiness, severity and the attack--fault trade-off}

\begin{figure}[t]
\centering
\includegraphics[width=\columnwidth]{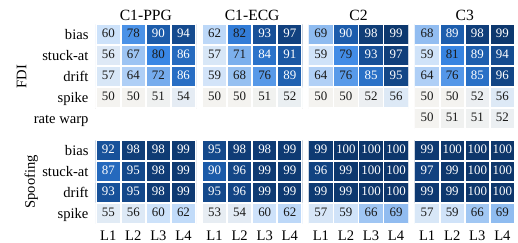}
\caption{Host-matched AUROC ($\times100$; 50 = chance) by morphology and severity level, averaged over the six learned detectors and five seeds. Levels are ordinal within a morphology; the rate-warp levels are the factors 1.05--1.35. Replay has no levels and scores 50--52.}
\label{fig:grid}
\end{figure}

Detection follows the expected progression from replay to FDI to spoofing attacks (Table~\ref{tab:rq2}). At a validation FAR budget of 10\%, detectors identify 89--98\% of sustained spoofing attacks but only 38--68\% of sustained in-bound FDI attacks. Fig.~\ref{fig:grid} further breaks down results by morphology and severity level for each attacker setting.
Two attack families consistently defeat all six learned detectors.
\emph{Replay} yields AUPRC values of 0.076--0.110 against a 0.065 chance floor and AUROC values of 0.520--0.643, with overlapping confidence intervals across methods. At the matched operating budget, no detector identifies more than 20.2\% of replay attacks. Because replayed segments are genuine physiological signals, reconstruction, forecasting, and likelihood-based detectors have little evidence to exploit. The physiological-coherence baseline specifically targets the ECG-PPG timing mismatch introduced by replay but reaches only AUPRC 0.089 (0.077--0.110) and AUROC 0.614. In 10\,s ICU windows, beat-timing variability is simply too large for a 2\,s replay to stand out. Standard beat detectors from NeuroKit2 and Elgendi's PPG method~\cite{makowski2021,elgendi2013} produce a similar result (replay AUROC 0.582; host-matched AUROC 0.549).

\emph{Short spikes.} FDI-spike attacks achieve AUPRC values of only 0.142--0.201 against a 0.122 chance baseline, while L1 FDI-spike attacks reach 0.041--0.054 against a 0.033 floor. Spoofed spikes are only slightly easier to detect (Fig.~\ref{fig:grid}), suggesting that their short duration (3--5 samples) is the primary challenge, not their amplitude. Removing $10\times$ temporal pooling from LSTM-AE improves replay and FDI-spike AUPRC by similar amounts (0.066 and 0.064, respectively), indicating a general benefit of higher temporal resolution. However, USAD, which already processes full-resolution windows, performs no better on spikes (AUPRC 0.142).

\begin{table*}[t]
\centering
\caption{Attack subsets, C1-PPG (five seeds). Left: threshold-free AUPRC of each attack subset against clean normals, with the subset's chance floor in parentheses, and threshold-free AUROC of fault and event windows against normals. Right: detection with thresholds at 10\% FAR on validation normals, applied unchanged to test (deployment view), and fault FAR at an equal 5\% FAR set on test normals.}
\label{tab:rq2}
\footnotesize
\setlength{\tabcolsep}{2pt}
\begin{tabular}{@{}l cccccc ccc c cccccc cc@{}}
\toprule
& \multicolumn{6}{c}{Threshold-free AUPRC / AUROC (attacks)} & \multicolumn{3}{c}{AUROC vs.\ normal} & & \multicolumn{6}{c}{At 10\% validation FAR} & \multicolumn{2}{c}{Equal 5\% FAR}\\
\cmidrule(lr){2-7}\cmidrule(lr){8-10}\cmidrule(lr){12-17}\cmidrule(lr){18-19}
Detector & \makecell{Replay\\(.065)} & \makecell{Replay\\AUROC} & \makecell{Spike\\(.122)} & \makecell{Spike\\L1 (.033)} & \makecell{FDI\\sust.\ (.293)} & \makecell{Spoof.\\sust.\ (.293)} & \makecell{Artif.} & \makecell{Mon.\\FA} & Event & & \makecell{FAR\\normal} & \makecell{FAR\\fault} & Ratio & Replay & \makecell{FDI\\sust.} & \makecell{Spoof.\\sust.} & \makecell{FAR\\artif.} & \makecell{FAR\\mon.\ FA}\\
\midrule
Isolation Forest & .110 & .643 & .201 & .054 & .743 & .964 & .661 & .556 & .552 && .050 & .263 & 5.26 & .099 & .446 & .961 & .290 & .179\\
OmniAnomaly & .103 & .604 & .171 & .053 & .658 & .930 & .351 & .547 & .612 && .106 & .097 & 0.92 & .202 & .501 & .887 & .037 & .161\\
GDN & .098 & .568 & .169 & .051 & .579 & .938 & .367 & .544 & .687 && .082 & .150 & 1.83 & .180 & .420 & .945 & .089 & .207\\
LSTM-AE & .088 & .532 & .162 & .049 & .678 & .977 & .392 & .538 & .557 && .070 & .128 & 1.83 & .165 & .506 & .972 & .070 & .189\\
TranAD & .084 & .531 & .153 & .044 & .759 & .970 & .431 & .548 & .588 && .082 & .134 & 1.63 & .148 & .680 & .981 & .081 & .185\\
USAD & .076 & .520 & .142 & .041 & .562 & .940 & .450 & .538 & .530 && .088 & .124 & 1.41 & .121 & .381 & .909 & .066 & .154\\
\midrule
Coherence & .089 & .614 & .150 & .042 & .381 & .480 & .609 & .476 & .636 && .056 & .244 & 4.36 & .072 & .099 & .161 & .285 & .091\\
\bottomrule
\end{tabular}
\end{table*}

\emph{Coupling features and faults.} The features that most improve detection of difficult attacks also increase sensitivity to physiological faults. Isolation Forest performs best on replay, FDI-spike, and L1 FDI-spike attacks and, at the matched budget, flags 26.3\% of fault windows while producing the lowest FAR on normal windows (5.0\%). Because this low FAR inflates the fault-to-normal ratio (5.3 versus 0.9--1.8 for other methods), we examine two corrected comparisons. Without thresholding, artifact-confirmed windows achieve AUROC 0.661 (0.502--0.833) against normals for Isolation Forest and 0.609 for the coherence baseline. In contrast, the five deep models score artifact windows below normals (AUROC 0.35--0.45), largely because flat or clipped signals are easy to reconstruct. Monitor false alarms remain near chance for all detectors (0.48--0.56). A detached lead disrupts ECG--PPG coupling more directly than a monitor false alarm. At a common 5\% FAR, Isolation Forest still detects 29.0\% of artifact-confirmed windows, compared with only 3.7--8.9\% for the deep models, confirming that the finding is not an artifact of threshold selection.
Genuine physiological events remain difficult to distinguish from normal windows (AUROC 0.53--0.69), although performance varies by alarm type. Ventricular flutter/fibrillation achieves the highest scores (up to 0.89, although only six test examples exist), whereas bradycardia and ventricular tachycardia can score below normal windows (AUROC as low as 0.33 and 0.38).

\subsection{RQ3: Attacker capability}

Table~\ref{tab:cap} evaluates all four attacker settings using a stricter threshold corresponding to 5\% FAR on validation normals and reports host-matched AUROC across five seeds. Host-matched AUROC is consistently 0.02--0.05 lower than AUROC calculated against all test normals, indicating that some separation in Table~\ref{tab:rq1} arises from host selection.

The \emph{attacked channel} has little influence on performance. Applying the same 34 attack conditions to ECG changes host-matched AUROC by only $-0.02$ to $+0.06$ relative to PPG, depending on the detector, with no consistent trend. Thus, the findings are not specific to attacks on PPG.

\emph{Independent two-channel attacks} are the easiest to detect. Under C2, host-matched AUROC increases for every learned detector by 0.05--0.08 relative to C1-PPG, reaching 0.78--0.89.

\emph{Coordinated attacks} remain the most challenging. Across all C3 conditions, host-matched AUROC values (0.74--0.85) lie between those of C1 and C2 because coordinated FDI and spoofing attacks still move both channels away from normal physiology. However, the two consistency-preserving C3 families operate at chance. Joint replay achieves host-matched AUROC values of 0.50--0.55 (Isolation Forest: 0.530, 0.509--0.553), essentially identical to single-channel replay. Similarly, the rate warp achieves AUROC values of 0.49--0.54 and is flagged in only 4.1--9.7\% of cases, close to the 1.5--3.9\% FAR observed on clean windows. Because rate warps modify ECG and PPG together, cross-channel features remain normal and the deep models observe a plausible physiological rhythm. The coherence baseline also remains near chance (Table~\ref{tab:cap}). Although the warp changes pulse-transit time, the effect is too small to be reliably measured with current beat-detection accuracy.

\begin{table*}[t]
\centering
\caption{Attacker capability (threshold at 5\% FAR on validation normals). Host-matched AUROC compares each attacked window with its own clean host among the 188 test hosts (chance 0.5). Det.\ = share of attacks above the threshold; test FAR = achieved rate on all clean test windows. Means over five seeds (sd $\le$0.03); the coherence baseline is deterministic.}
\label{tab:cap}
\footnotesize
\setlength{\tabcolsep}{5pt}
\begin{tabular}{@{}l cccc ccc cc@{}}
\toprule
& \multicolumn{4}{c}{Host-matched AUROC, all conditions} & \multicolumn{3}{c}{Host-matched AUROC, hardest families} & \multicolumn{2}{c}{At 5\% validation FAR}\\
\cmidrule(lr){2-5}\cmidrule(lr){6-8}\cmidrule(lr){9-10}
Detector & C1-PPG & C1-ECG & C2 & C3 & C1-PPG replay & C3 joint replay & C3 rate warp & Det.\ rate warp & Test FAR\\
\midrule
Isolation Forest & .827 & .809 & .890 & .845 & .533 & .530 & .541 & .041 & .015\\
USAD & .726 & .731 & .779 & .738 & .499 & .496 & .491 & .060 & .038\\
LSTM-AE & .754 & .761 & .805 & .760 & .494 & .499 & .488 & .080 & .034\\
GDN & .724 & .782 & .807 & .765 & .519 & .525 & .519 & .093 & .031\\
TranAD & .778 & .800 & .829 & .786 & .494 & .503 & .494 & .078 & .036\\
OmniAnomaly & .745 & .746 & .812 & .777 & .550 & .553 & .535 & .097 & .039\\
\midrule
Coherence & .598 & .585 & .580 & .541 & .560 & .548 & .535 & .031 & .021\\
\midrule
Test attack windows & 6,392 & 6,392 & 6,204 & 7,144 & 376 & 376 & 752 & 752 & --\\
\bottomrule
\end{tabular}
\end{table*}

\subsection{RQ4: Alarm-level counterfactuals}

\emph{Validation.} We first evaluate the reference monitor on the window ending at the alarm time, which contains the annotated event onset. The ECG-only monitor, $g_{\mathrm{ECG}}$, fires on 88\%, 85\%, and 82\% of true asystole, bradycardia, and tachycardia alarms, respectively. It also fires on 52\%, 42\%, and 40\% of corresponding false alarms, reflecting the fact that the bedside monitor alarmed as well, and on 15\% of pre-alarm normal windows. Requiring agreement between ECG and PPG ($g_{\mathrm{fused}}$) substantially reduces these false alarms, lowering rates to 4.5\%, 0\%, and 20\% (1 of 5 cases), and reducing the normal-window alarm rate to 3.7\%. This improvement comes at the cost of lower sensitivity, which falls to 65\%, 56\%, and 20\%, respectively, because pulse detection from PPG is least reliable at high heart rates. Although the monitor is imperfect, counterfactual analysis requires a fixed and reproducible decision rule.

\emph{Alarm creation and suppression.} Table~\ref{tab:alarm} compares the output of $g$ on each attacked window with its clean counterpart. An ECG-only attacker can create alarms in $g_{\mathrm{ECG}}$, affecting 17\% of FDI windows and 41\% of spoofing windows, but creates almost none in $g_{\mathrm{fused}}$ (at most 2.2\%). A PPG-only attacker cannot influence $g_{\mathrm{ECG}}$ and generates few fused alarms (at most 5.5\%). However, the same attacker suppresses 23--45\% of fused alarms that would otherwise occur, including 38\% suppression under a 2\,s replay. The agreement rule that effectively removes false alarms therefore gives a single-channel attacker the ability to veto alarms.
Two-channel attacks affect both monitors. Spoofing attacks under C2 and C3 create alarms in approximately 37\% of affected windows for $g_{\mathrm{fused}}$. Coordinated attacks that evade detection also alter clinical decisions. Joint replay attacks, which remain undetected in RQ3, suppress 48\% of fused alarms, while the coordinated rate-warp attack suppresses 24\%. Replay, the hardest attack family to detect, suppresses alarms almost as often as obvious out-of-range spoofing attacks (26--48\% versus 28--45\% of fused alarms across attacker settings). In contrast, spike attacks, which are also difficult to detect, cause little practical harm. They create at most 2.1\% of alarms and suppress only 2--17\% of fused alarms because modifying or removing a single beat rarely changes a decision based on 5- or 17-beat criteria.

\begin{table}[t]
\centering
\caption{Alarm-level counterfactuals: share of attacked windows whose rate-alarm decision differs from that of the clean host. Creation on the 848 ($g_{\mathrm{ECG}}$) and 1,093 ($g_{\mathrm{fused}}$) pre-alarm hosts without a clean alarm; suppression on the 125 and 47 post-alarm hosts of true-alarm records with one.}
\label{tab:alarm}
\footnotesize
\setlength{\tabcolsep}{4pt}
\begin{tabular}{@{}ll cc cc@{}}
\toprule
& & \multicolumn{2}{c}{Creation} & \multicolumn{2}{c}{Suppression}\\
\cmidrule(lr){3-4}\cmidrule(lr){5-6}
Setting & Family & $g_{\mathrm{ECG}}$ & $g_{\mathrm{fused}}$ & $g_{\mathrm{ECG}}$ & $g_{\mathrm{fused}}$\\
\midrule
C1-PPG & replay & 0 & .006 & 0 & .383\\
 & FDI & 0 & .023 & 0 & .229\\
 & spoofing & 0 & .055 & 0 & .447\\
C1-ECG & replay & .060 & .002 & .360 & .255\\
 & FDI & .167 & .010 & .101 & .162\\
 & spoofing & .407 & .022 & .228 & .396\\
C2 & replay & .021 & .005 & .232 & .362\\
 & FDI & .167 & .139 & .116 & .188\\
 & spoofing & .408 & .372 & .218 & .299\\
C3 & joint replay & .057 & .007 & .324 & .479\\
 & rate warp & .075 & .031 & .190 & .239\\
 & FDI & .168 & .143 & .116 & .168\\
 & spoofing & .403 & .367 & .234 & .283\\
\bottomrule
\end{tabular}
\end{table}

\subsection{RQ5: Cause discrimination}

We evaluate a random forest trained on the training and validation records to distinguish among event, fault, and attack, which reflects the decision problem clinicians face after an alarm occurs. The classifier recognizes attacks well (F1 = 0.967), faults moderately (F1 = 0.674), and genuine physiological events very poorly (F1 = 0.040). Only 2 of 92 event windows are correctly identified; 71 (77.2\%) are classified as attacks and 19 (20.7\%) as faults. No alarm category is reliably recognized. By alarm type, the pattern is similar: both asystole examples are classified as faults, while 61--83\% of windows from every other event type are labeled as attacks and the remainder as faults. Conversely, 40.2\% of fault windows are predicted as attacks, consistent with the ambiguity between faults and attacks observed in RQ2.

Table~\ref{tab:rq4v} investigates whether this behavior persists after controlling for position, labeling, and class imbalance. Attack detection remains unchanged (F1 $\approx$ 0.97 across all four attacker settings), indicating that the strong attack performance is not driven by positional shortcuts. Event recognition, however, remains extremely poor (F1 0.00--0.02), while fault F1 drops to 0.08--0.21. For C1-PPG, 87 of 92 post-alarm event windows and 159 of 184 post-alarm fault windows are classified as attacks. Restricting the fault class to artifact-confirmed windows gives artifact F1 = 0.58, but 130 of the 164 held-out monitor-false-alarm windows are still classified as attacks. When monitor false alarms are introduced as a separate class, they are never recognized (F1 = 0.00). \emph{Re-slicing} post-alarm events at a 1\,s stride inside the training folds raises event F1 from 0.00 to 0.22 (0.27 without class weights), with the test distribution unchanged, and relabeling alarm-zone windows by verdict gives 0.20. Thus, the difficulty in recognizing genuine physiological events is not explained by window position or label construction, and additional event data provides only limited improvement.

\begin{table}[t]
\centering
\caption{Discrimination under controls (default forest: 300 trees, class-balanced): per-class F1, event F1 with its 95\% record-grouped interval, and macro-F1. The rerun recognizes 0 of 92 events where the run reported in the text recognizes 2, because of sample order. Four classes: fault = artifact-confirmed; monitor false alarms reach F1 0.}
\label{tab:rq4v}
\footnotesize
\setlength{\tabcolsep}{2.2pt}
\begin{tabular}{@{}lcccc@{}}
\toprule
Variant & Event & Fault & Attack & Macro\\
\midrule
Default setting (rerun) & .000 [.00, .00] & .668 & .967 & .545\\
Artifact-confirmed fault only & .039 [.00, .12] & .583 & .970 & .531\\
Four classes & .020 [.00, .07] & .538 & .958 & .379\\
Position-matched, C1-PPG & .000 [.00, .00] & .206 & .972 & .393\\
Position-matched, C1-ECG & .000 [.00, .00] & .093 & .971 & .354\\
Position-matched, C2 & .021 [.00, .07] & .204 & .971 & .399\\
Position-matched, C3 & .000 [.00, .00] & .081 & .973 & .351\\
Events re-sliced at 1\,s & .222 [.12, .33] & .664 & .966 & .617\\
\quad without class weights & .272 [.14, .40] & .669 & .969 & .637\\
Alarm zone relabeled & .202 [.10, .30] & .690 & .962 & .618\\
\bottomrule
\end{tabular}
\end{table}

\subsection{Attribution, label and leakage audits}
\label{sec:audit}

\emph{Channel attribution.} The addition of C1-ECG attacks creates a nontrivial attribution problem because either channel may be manipulated. A baseline that always predicts PPG would achieve only 50\% accuracy. We therefore evaluate a per-channel Isolation Forest that attributes the attack to the channel with the larger anomaly score. On C1 test attacks, it achieves a top-1 attribution accuracy of 0.863, including 0.537 for replay attacks, 0.822 for FDI attacks, and 0.946 for spoofing attacks. Performance is similar for ECG targets (0.851) and PPG targets (0.875). Replay again operates near chance, so attribution on the benchmark is a nontrivial task; dedicated causal attribution methods remain future work.

\emph{Alarm-zone labels.} Removing alarm-zone windows from the normal class increases C1-PPG AUPRC by 0.005--0.008 and replay AUPRC by 0.004--0.007 for all learned detectors without changing their ranking.

\emph{Leakage (WUSTL-EHMS-2020).} Although WUSTL-EHMS-2020 is not part of the benchmark because its network traffic is not synchronized with the CinC waveforms, it provides an independent public IoMT dataset for evaluating data leakage. Because all data come from a single session, we use stratified 70/30 random splits across five seeds, a setting where identifier leakage is likely. Table~\ref{tab:wustl} removes groups of features in turn. Host identifiers, particularly source MAC addresses and ports, account for most of the near-perfect classification performance. Source MAC alone contributes 59\% of total feature importance, while identifiers combined with TCP flags contribute 66\%. Unlike identifiers, TCP flags can contain legitimate behavioral information and improve Macro-F1 and AUPRC by approximately 0.03. The running packet counter contributes an additional 0.06 AUPRC because, in a random split from a single session, it effectively encodes time and reveals attack periods. Under our evaluation protocol, near-perfect results depend heavily on identifier leakage, and we do not reproduce the settings used in earlier studies.

\begin{table}[t]
\centering
\caption{WUSTL-EHMS-2020 leakage audit: class-balanced random forest (200 trees), mean over five stratified 70/30 splits (sd $\le$0.02).}
\label{tab:wustl}
\footnotesize
\setlength{\tabcolsep}{3pt}
\begin{tabular}{@{}lccc@{}}
\toprule
Features & \makecell{Attack\\recall} & Macro-F1 & AUPRC\\
\midrule
All (identifiers kept) & 1.000 & 1.000 & 1.000\\
No identifiers, no flags & .494 & .808 & .790\\
No identifiers, flags kept & .554 & .838 & .824\\
No identifiers, flags or packet counter & .471 & .796 & .734\\
Network only (no identifiers, no flags) & .462 & .788 & .680\\
Biometrics only & .648 & .734 & .655\\
\bottomrule
\end{tabular}
\end{table}

\section{Discussion}
\label{sec:disc}

\textbf{Common failure mode.} The difficulty with replay attacks and short spikes should be interpreted as a limitation of current general-purpose detectors, not an inherent limit of attack detection. The physiological-coherence baseline does not solve the problem either and remains ineffective against amplitude-modifying attacks that preserve beat timing (Table~\ref{tab:rq2}).

\textbf{What physiological signals alone can detect.} For C1 and C2 attacks, inconsistencies between ECG and PPG provide useful evidence, although they cannot distinguish malicious manipulation from faults such as a detached lead. A C3 attacker deliberately preserves cross-channel consistency and therefore removes this source of evidence. As a result, coordinated replay and rate-warp attacks appear physiologically plausible to every detector evaluated. Detecting such attacks likely requires information beyond the waveform itself, including temporal context across windows, device provenance, cryptographic authentication, or network-level evidence.

\textbf{Limitations.} (i) Attacks are semi-synthetic and predefined, not collected from real deployments. (ii) The alarm oracle covers only the three rate-based CinC alarms and uses a reference implementation, not a commercial monitor. (iii) The benchmark does not include synchronized network traffic. (iv) All physiological data come from 5--5.5\,min alarm-centered CinC records, and splits are record-disjoint, with patient-disjointness unverified. (v) For six of the sixteen FDI cells (Table~\ref{tab:sev}), in-bound perturbation magnitudes reach the plausibility boundary, making L3--L4 severity levels ordinal instead of proportional. (vi) Detector hyperparameters were fixed and not tuned, and among pooled deep models only LSTM-AE was evaluated without temporal pooling. (vii) Both the coherence baseline and alarm oracle depend on beat detection within 10\,s windows. (viii) Because drift, spike, and stuck-at FDI attacks intentionally resemble sensor faults, some fault--attack confusion observed in RQ5 reflects the threat model itself and does not necessarily indicate classifier error.

\section{Conclusion}

IoMT-SecAlarmBench is a semi-synthetic benchmark for integrity attacks on authentic coupled ECG and PPG recordings. The benchmark preserves the clean waveform underlying every injected attack, converts each clean--attacked pair into an alarm-level counterfactual using a specified rate-alarm monitor, and includes faults and physiological events drawn from the same records. The baseline evaluations show that a coordinated attacker can perform consistency-preserving replay and rate-warp attacks that evade every evaluated detector while suppressing alarms that the clean signal would have triggered.
Future work will extend the benchmark with MIMIC-III recordings~\cite{moody2020}, synchronized network traces, causal attribution methods, and adaptive attack models. 

\noindent We release the benchmark code, attack generators, preprocessing pipeline, evaluation framework, and complete datasheet at \url{https://github.com/techosystem/IoMT_SecAlarmBench}.

\end{document}